\documentclass{PoS}
\newcommand{\be}{\begin{equation}}
\newcommand{\ee}{\end{equation}}

\def\be{\begin{equation}}
\def\ee{\end{equation}}
\def\beq{\begin{eqnarray}}
\def\eeq{\end{eqnarray}}

\def\({\left (}
\def\){\right )}
\def\[{\left [}
\def\[{\right ]}

\def\ba{\begin{eqnarray}}
\def\ea{\end{eqnarray}}

\title{Non-linear evolution equation for gluon density at large values of coupling constant}

\ShortTitle{Nonlinear evolution}

\author{\speaker{Krzysztof Kutak}\\
       Institute for Nuclear Physics, Polish Academy of Science\\
Radzikowskiego 152, Krakow, Polska.\\
        E-mail: \email{krzysztof.kutak@ifj.edu.pl}}

\abstract{The recently proposed nonlinear evolution equation \cite{Kutak:2013hda} for unintegrated gluon densities valid for large values of the QCD coupling constant $\bar{\alpha} _s$ is presented.
In particular we outline its derivation, numerical solution and obtained saturation scale.}

\FullConference{XXII. International Workshop on Deep-Inelastic Scattering and Related Subjects,\\
		28 April - 2 May 2014\\
		Warsaw, Poland}

\begin{document}

\section{Introduction}
A strongly coupled quantum field theory like quantum chromodynamics is the basic framework which is used in the interpretation of hadronic observables data from the high-energy physics experiments. Despite its mature status many open questions remain, and the full theoretical description is far from complete, as neither perturbative methods nor lattice gauge theory can provide a full description of hadronic phenomena. One of the open problems is, for instance, gluon saturation \cite{Gribov:1984tu}, which is expected on theoretical grounds. Another open problem is the derivation of the dynamics of strongly coupled systems, such as quark-gluon plasma, directly from a QCD Lagrangian.
In our approach to bridge weak and strong coupling dynamics at least for some quantities of intrest like gluon density we use an appropriately resummed nonlinear BFKL equation and extend it to the nonlinear regime. We solve the equation and calulate a saturation scale it generates.\\

\section{The BFKL equation in diffusion approximation} \label{BFKLrevised}
The BFKL equation written for the unintegrated gluon density in the integral form reads
\be
f(x,k^2)=f_0(x,k^2)+\bar{\alpha}_sk^2\int_{x/x_0}^1\frac{dz}{z}\int_0^\infty\frac{dl^2}{l^2}\left[\frac{f(x/z,l^2)-f(x/z,k^2)}{|l^2-k^2|}+\frac{f(x/z,k^2)}{\sqrt{4l^4+k^4}}\right],
\label{eq:BFKL1}
\ee
where $x$ is a longitudinal momentum fraction carried by the gluon, $k$ is the modulus of its transversal momentum. The LO BFKL equation can be solved by the Mellin transform. The characterisctic function reads:
$\chi(\gamma)=2\psi(1)-\psi(1-\gamma)-\psi(\gamma)$, where $\psi$ is a digamma function. The solution to Eq. (\ref{eq:BFKL1}) can be written as
\be
f(x,k^2)=\frac{1}{2\pi i}\int d\gamma (k^2)^{\gamma}\frac{1}{2\pi i}\int d\omega x^{-\omega}\frac{\omega {\overline f}_0(\omega,\gamma)}{\omega-\overline{\alpha_s}\chi(\gamma)}.
\ee
In order to evaluate the integral above one needs to know the characteristic function along the imaginary axis in the $\gamma$ plane.
Knowing that the characteristic function has a saddle point along the $\gamma= 1/2+i\nu$ contour, we can write the solution as
\be
{\cal F}(x,k^2)={\cal F}(x_0,1/2)\frac{1}{\sqrt{4\pi\ln(x_0/x)1/2\lambda^{\prime}}}e^{\lambda\ln(x_0/x)-1/2\ln(k^2/k_0^2)}e^{\frac{-\ln(k^2/k_0^2) ^2}{4\,1/2\lambda^{\prime}\ln(x_0/x)}},
\ee
where ${\cal F}(x_0,1/2)=f(x_0,1/2)/k^2$ and $\chi(1/2+i\nu)\approx\lambda-\frac{1}{2} \lambda^{\prime}\nu^2$ with $\lambda=\bar{\alpha}_s4\ln2$ and $\lambda^{\prime}=\bar{\alpha}_s\zeta(3)$.
From this explicit form, one may extract the coefficients of the diffusion equation
\be \label{diffusioneq}
\partial_Y{\cal F}(Y,\rho)=\frac{1}{2} \lambda^{\prime} \partial^2_\rho {\cal F}(Y,\rho)+\frac{1}{2}\lambda^{\prime}\partial_\rho{\cal F}(Y,\rho)+(\lambda+\lambda^{\prime}/8){\cal F}(Y,\rho).
\ee
where ${\cal F}(x,k^2)\equiv f(x,k^2)/k^2$ and $Y=\ln \frac{x_0}{x}\,\,\rho = \ln \frac{k^2}{k_0 ^2}$.

\section{The BFKL equation with higher order corrections and the gluon density in the whole range of coupling constant}\label{BFKLstrong}
The BFKL equation has been obtained at NLO accuracy in Refs. \cite{Fadin:1998py,Balitsky:2008zza}. However, in order to make the eigenvalue of the kernel stable, one needs to perform resummations of corrections to infinite order. One source of such corrections is provided by the so-called kinematical constraint effects. 
\begin{figure}[t!]
  \begin{picture}(30,30)
   \put(30, -110){
      \includegraphics{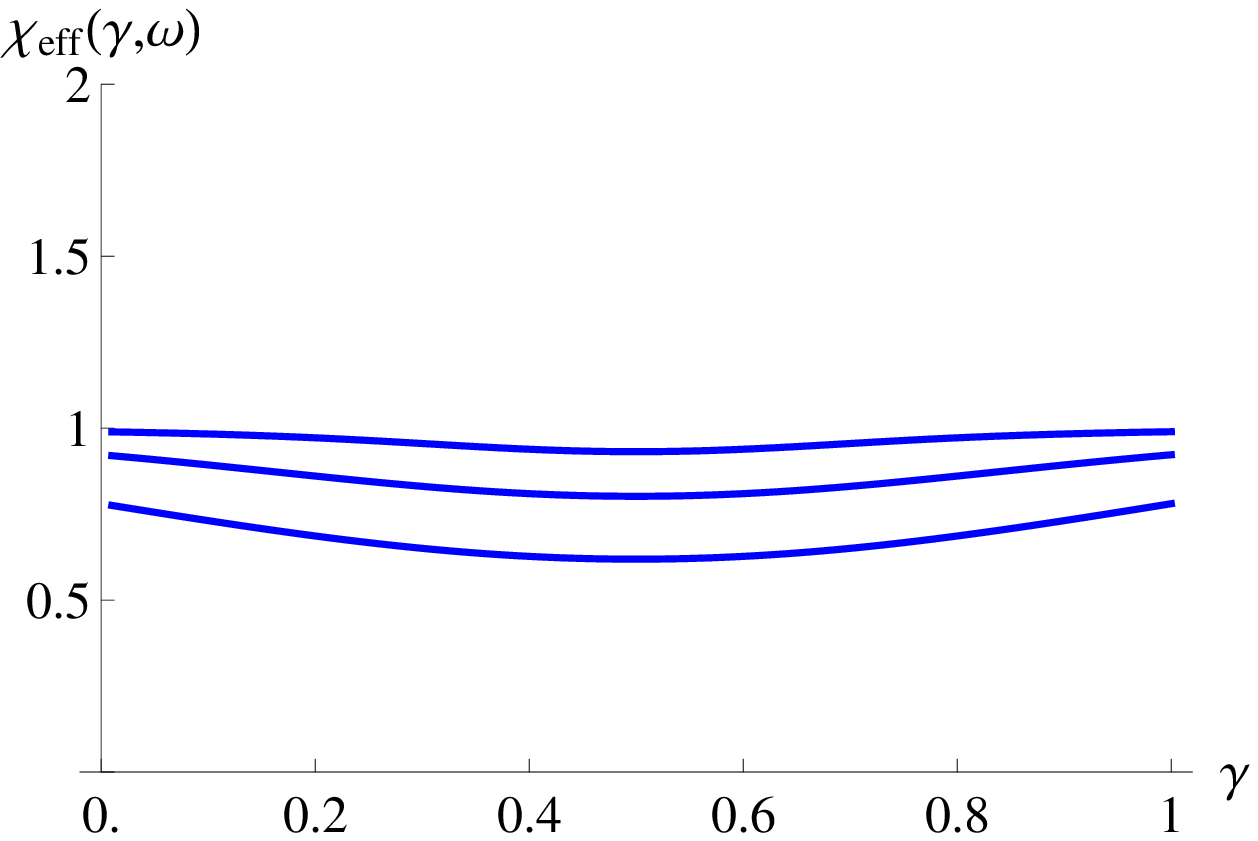}
    }

\put(280, -110){
      \includegraphics{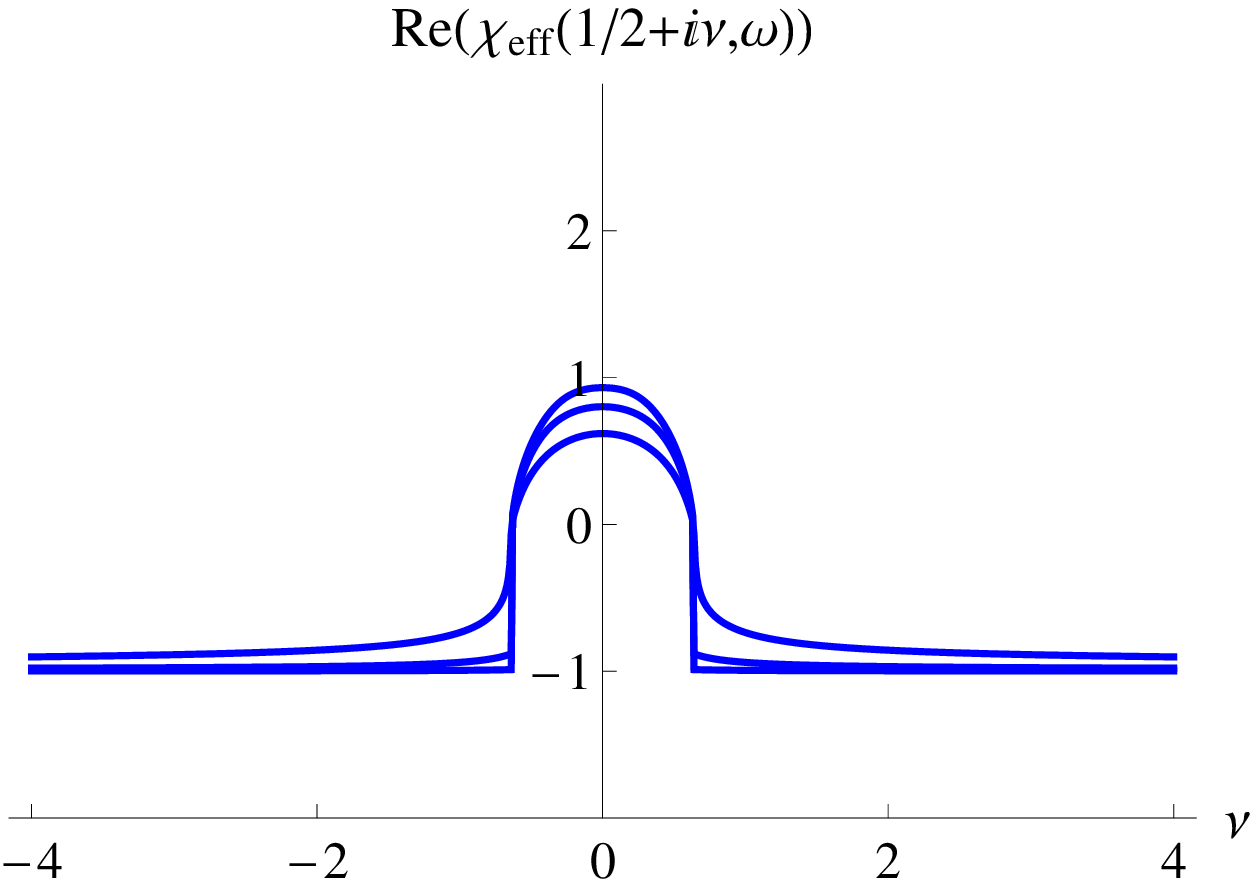}
    }

\end{picture}
\vspace{4cm}
\caption{\em \small  Kinematical constraint effects and resummation effects. Upper right plot: Function $\chi_{eff}(\gamma,\omega)$ along the real contour for $\bar{\alpha}_s=0.2,0.5,1,2$.
 Upper left plot: Function $\chi_{eff}(\gamma,\omega)$ along the imaginary contour for $\bar{\alpha}_s=0.2,0.5,1,2$. Lower left plot:  Function $\chi_{eff}(\gamma,\omega)$ along the real contour for $\bar{\alpha}_s=2,10,100$.  Lower right plot:  Function $\chi_{eff}(\gamma,\omega)$ along the real contour for $\bar{\alpha}_s=2,10,100$.}
\vspace{1.5cm}
\label{fig:kinres}
\end{figure}
It has been suggested in Ref. \cite{Stasto:2007uv} that
\be
\frac{1}{{\bar \alpha}_s}=\gamma^{(0)}(\omega)\chi_{k.c.}(\gamma,\omega),
\label{eq:effeigen}
\ee
where
\be
\gamma^{(0)}(\omega)=\frac{1}{\omega}+A(\omega)
\ee
is the LO DGLAP anomalous dimension and
\be
A(\omega)=-\frac{1}{\omega+1}+\frac{1}{\omega+2}-\frac{1}{\omega+3}-\psi(2+\omega)+\psi(1)+\frac{11}{12} 
\ee
while the kinematical constraind improved $\chi$ function reads:
\be
\chi_{k.c.}=2\psi(1)-\psi(1-\gamma+\omega/2)-\psi(\gamma+\omega/2)
\ee
\begin{figure}[t!]
  \begin{picture}(30,30)
    \put(0, -110){
      \includegraphics{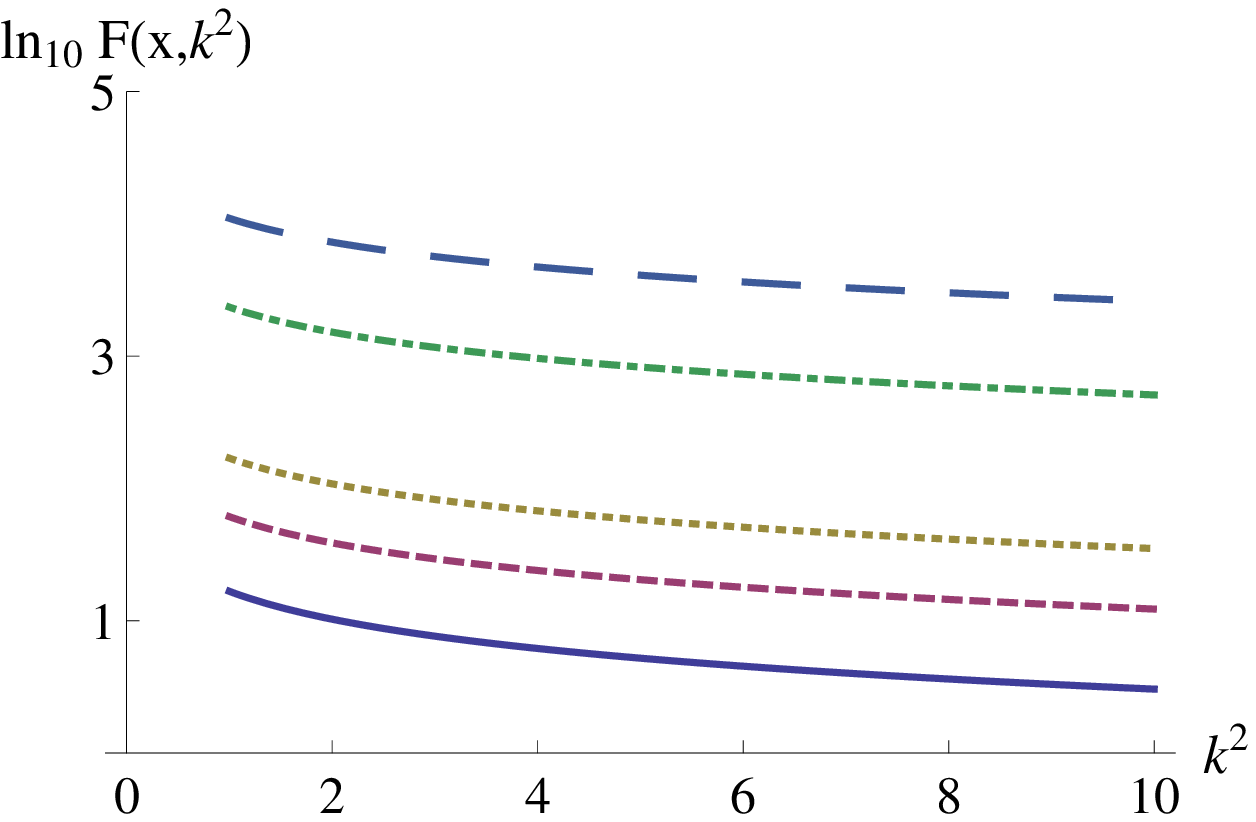}
    }

 \put(240, -110){
      \includegraphics{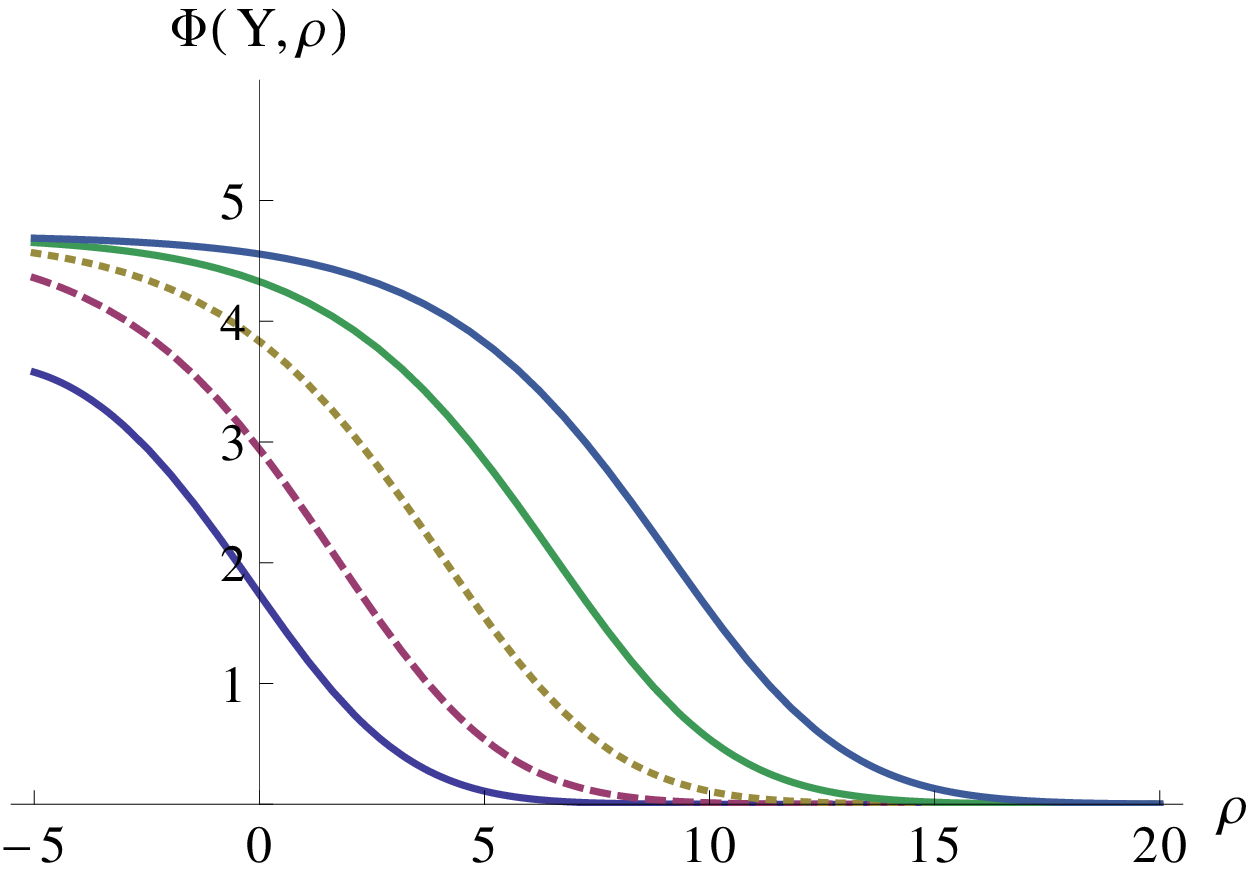}
    }
\end{picture}
\vspace{4cm}
\caption{\em \small  Gluon density obtained for various values of the coupling constant ${\bar\alpha}_s\!=\!0.2,\,0.5,\,1,\,10, \,10^4$. The densities decrease for a smaller coupling constant. Left, $x=10^{-6}$; right, $x=10^{-4}$. }
\vspace{1cm}
\label{fig:plotvel66}
\end{figure}
Equation (\ref{eq:effeigen}) provides a resummation of DGLAP gluon anomalous dimension at LO (missing in BFKL) and kinematical effects. It can be written as an effective eigenvalue equation in the form
$
\omega= \chi_{eff}(\gamma,\omega),
$
with
$
\chi_{eff}(\gamma,\omega)=\bar{\alpha}_s\chi_{k.c.}(\gamma,\omega)\left(1+A\omega\right).
$
The model introduced above has beeen studied in the context of large strong coupling limit. The crucial behavior, providing energy conservation, is the vanishing of the eigenvalue function when $\omega\rightarrow 1$ see Fig \ref{fig:kinres}.
As a practical application of the result obtained in Ref. \cite{Stasto:2007uv}, we ask a question about the  properties of the gluon density while evaluated at the increasing values of strong coupling.
Naively, one can think that if the coupling constant increases, the number of gluons vanishes or at least it is constant. Below, we show that this is not the case, the gluon density grows, and we get infinitely many soft gluons. 
The resulting gluon density obtained from strongly coupled BFKL for different values of the coupling constant is shown on left panel of Fig. \ref{fig:plotvel66}.
The form of the eigenvalue function at strong values of coupling constant is:
\be
\chi_{eff\,\infty}(\omega,1/2+i\nu)=1.02795\, -2.04635 \nu ^2\equiv\lambda_{st}-\frac{1}{2}\lambda_{st}^{\prime}\nu^2,
\label{eq:fitquad}
\ee
Formula (\ref{eq:fitquad}) can be used to obtain analytically a solution of the BFKL equation in the strong coupling regime and to deduce a partial differential equation which it obeys:
\be
\partial_Y\Phi(Y,\rho)=\frac{1}{2}\lambda_{st}^{\prime} \partial^2_\rho \Phi(Y,\rho)+\frac{1}{2}\lambda_{st}^{\prime}\partial_\rho\Phi(Y,\rho)+(\lambda_{st}+\lambda_{st}^{\prime}/8)\Phi(Y,\rho),
\label{eq:BFKLstrong}
\ee
where the values are read off from formula (\ref{eq:fitquad}) $\lambda_{st}^{\prime}=4.08, \,\lambda_{st}=1.02$

\section{The BK equation in the limit of infinite coupling constant $\bar{\alpha} _s$}\label{nonlineraeq}
The linear BFKL evolution equation misses a very important aspect of the high-energy scattering, namely, the saturation physics.  As pointed out in the introduction, several approaches were constructed in order to include non-linear effects, like multiple scattering and gluon saturation, responsible for the unitarization of the scattering amplitudes.  A particularly useful and simple enough approach to unitarize the cross section locally at given impact parameter is the Balitsky-Kovchegov (BK) equation,
which reads:
\be
\label{eq:BFKL}
\partial_{\ln1/x}\Phi_1(x,k^2)=\bar{\alpha}_s\int_0^\infty\frac{dl^2}{l^2}\left[\frac{l^2\Phi(x,l^2)-k^2\Phi(x,k^2)}{|l^2-k^2|}+\frac{k^2\Phi(x,k^2)}{\sqrt{4l^4+k^4}}\right]-\bar{\alpha}_s\frac{\overline\alpha_s}{\pi R^2}\Phi^2(x,k^2).
\ee
We note that, if one neglects the non-linear term, one recovers the linear BFKL equation.
The question arises how to extend the BK equation to strong coupling regime. In reference \cite{Kutak:2013hda} such an equation has been postulated based on numerical analysis and theoretical arguments:
\be
\partial_Y\Phi(Y,\rho)=\frac{1}{2}\lambda_{st}^{\prime} \partial^2_\rho \Phi(Y,\rho)+\frac{1}{2}\lambda_{st}^{\prime}\partial_\rho\Phi(Y,\rho)+(\lambda_{st}+\lambda_{st}^{\prime}/8)\Phi(Y,\rho)-\frac{\bar{\alpha}_s}{\pi R^2}\Phi^2(Y,\rho),
\ee
where
the values are read off from formula (\ref{eq:fitquad}): $\lambda_{st}^{\prime}=4.08,\,\lambda_{st}=1.02$. The equation has similar structure as the BK in diffusive approximation in the weak coupling limit  \cite{Munier:2003vc}.
The coefficient in front of the non-linear term has to be consistent with the large strong coupling limit we take in the linear part. We take the limit $\bar{\alpha}_s\rightarrow\infty$ ($\bar{\alpha}_s$ is essentially 't Hooft coupling) and assume a large target approximation ($R^2 \rightarrow \infty $), the ratio $\frac{\bar{\alpha}_s}{R^2}$ being fixed and we set it to unity.
The solution of the above equation
is presented on the right panel of Fig. \ref{fig:plotvel66}, and it shows that at some point where the shape of the curve flattens the number of gluons saturates.  
\begin{figure}[t!]
  \begin{picture}(30,30)

    \put(60, -110){
      \includegraphics{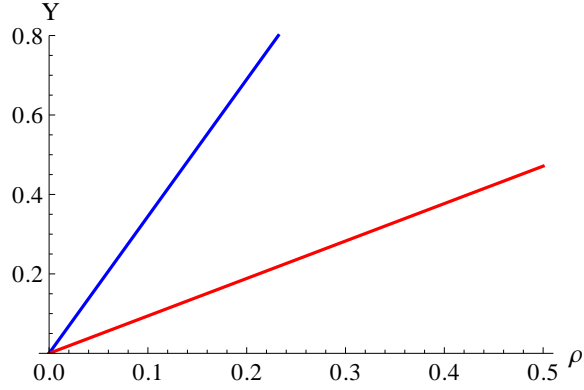}
    }

\end{picture}

\vspace{4cm}
\caption{\em {\small Red (lower) line: Saturation scale obtained from the solution of te new equation. Blue (upper) line: saturation scale as follows from the weak coupling equations and models: $Q_s^2(Y)\simeq e^{0.29\,Y}$}}.
\label{fig:plotvel7}
\end{figure}
The behavior of the gluon number density $\Phi(Y,\rho)$ is to be contrasted with the full form of the BK in the weak coupling regime, where the rate of production of gluons slows down but still diverges logarithmically and only approximately obeys the FKKP equation.

Now, we can calculate the saturation scale that follows from our equation and it reads $Q_{s}^2(Y)\simeq e^{1.06\,Y}$. For evaluation of saturation scale we use we use the method proposed in \cite{Kutak:2009zk}. We plot the strong coupling saturation scale together with the weak coupling counterpart in Fig \ref{fig:plotvel7}. The above result suggests that, at equal transversal momentum, saturation effects at strong coupling occur at smaller values of $\ln 1/x$ than at weak coupling. This can be easily understood, since the stronger the coupling is, the closer gluons are packed, and therefore the overlapping or screening takes at the initial values of evolution time . The result for saturation line is quite close to the one obtained in Refs. \cite{Hatta:2007he,Mueller:2008bt} by very different holographic methods, therefore pointing at the universality of the saturation phenomenon at large values of the coupling constant.

\section{Conclusions}
An approach to obtain an evolution equation that the gluon density obeys when the coupling constant is very large has been overviewed. The proposal is grounded in perturbative QCD framework that, through certain resummations, allows one to probe strong coupling physics. Solving this equation, we are able to extract the saturation scale, which agrees qualitatively with results from holography.

\section*{Acknowledgments}
This research has been supportrd by NCBiR Grant No. LIDER/02/35/L-2/10/NCBiR/2011.

\end{document}